\documentstyle[eqsecnum,aps]{revtex}
\newcommand{\sgn}{{\rm sgn}}

\newcommand{\veps}{\varepsilon}
\newcommand{\eps}{\epsilon}
\newcommand{\al}{\alpha}
\newcommand{\wa}{w_{\alpha}}

\newcommand{\KB}{K_{\rm B}}
\newcommand{\KF}{K_{\rm F}}

\newcommand{\go}{g_0}
\newcommand{\gr}{g_r}
\newcommand{\gq}{g_q}
\newcommand{\gl}{g_{1}}

\newcommand{\wo}{w_{0}}
\newcommand{\wl}{w_{1}}
\newcommand{\wB}{w_{\rm B}}
\newcommand{\wq}{w_{q}}
\newcommand{\wrkato}{w_{r}}
\newcommand{\wb}{w_{\beta}}
\newcommand{\wF}{w_{\rm F}}
\newcommand{\wFr}{w_{{\rm F},r}}
\newcommand{\wFq}{w_{{\rm F},q}}
\newcommand{\rhoo}{\rho_{0}}
\newcommand{\rhol}{\rho_{1}}
\newcommand{\rhoB}{\rho_{\rm B}}
\newcommand{\rhoF}{\rho_{\rm F}}
\newcommand{\rhoq}{\rho_{q}}
\newcommand{\rhor}{\rho_{r}}

\newcommand{\Mo}{M_0}

\newcommand{\B}{\rm B}
\newcommand{\F}{\rm F}
\newcommand{\PiB}{\Pi_{0}}
\newcommand{\PiF}{\Pi_{1}}
\newcommand{\Pio}{\Pi_{0}}
\newcommand{\Pil}{\Pi_{1}}
\newcommand{\Piq}{\Pi_{q}}
\newcommand{\Pir}{\Pi_{r}}
\newcommand{\po}{p_0}
\newcommand{\zetao}{\zeta_{0}}

\newcommand{\intp}{\int_{-\infty}^{\infty}dp}
\newcommand{\intk}{\int_{-\infty}^{\infty}dk}
\newcommand{\intkp}{\int_{-\infty}^{\infty}dk_1}
\newcommand{\intkpp}{\int_{-\infty}^{\infty}dk_2}
\newcommand{\GB}{G_{\rm 0}}
\newcommand{\GF}{K_{\F}}
\newcommand{\Gq}{G_q}
\newcommand{\Gr}{G_r}
\newcommand{\Go}{G_0}
\newcommand{\Gl}{K_{\F}}
\newcommand{\muB}{\mu_{\B}}
\newcommand{\muF}{\mu_{\F}}
\newcommand{\nuB}{\nu_{\B}}

\newcommand{\DeltaBF}{\Delta_{\rm BF}}
\newcommand{\DeltaFB}{\Delta_{\rm BF}^{-1}}
\begin{document}
\title{
Fractional Exclusion Statistics for the Multicomponent Sutherland Model} 
\author
{Yusuke {\sc KATO} and Yoshio {\sc KURAMOTO}}
\address{\it Department of Physics, Tohoku University, Sendai 980-77}
\maketitle
\begin{abstract}
We show by microscopic calculation that thermodynamics of the multicomponent Sutherland model is equivalent to that of a free particle system with fractional exclusion statistics at all temperatures. The parameters for exclusion statistics are given by the strength of the repulsive interaction, and have both intra- and inter-species components. 
We also show that low temperature properties of the system are described in terms of free fractional particles without the statistical parameters for different species. The effective exclusion statistics for intra-species at low temperatures depend on polarization of the system. 
\ \\
\end{abstract}
\section{Introduction}
State counting rule of a large number of identical particles depends on the quantum statistics; A number of bosons can occupy a single particle state ( no exclusion ) while the presence of a fermion in a state excludes other fermions (complete exclusion). Although particles and conventional elementary excitations obey bosonic or fermionic statistics so far, elementary excitations with fractional statistics attracted theoretical interests recently. Stimulated by the fractional quantum Hall systems and one-dimensional quantum spin chain, Haldane proposed a new type of quantum statistics from the viewpoint of state-counting. \cite{Haldanefrac} 
He defined a fractional generalization of the Pauli's exclusion principle so that it gives a state-counting rule interpolating the bosonic and fermionic one. This new type of quantum statistics is called the fractional exclusion statistics (FES). Subsequently, thermodynamics of single component free particles which obey FES was constructed, based on the new state-counting rule. \cite{Wu,Murphy} Thermodynamics was derived for the system of multicomponent particles obeying FES in ref. 4. All the above works start from the state-counting rule or entropy. Hence in refs. 1-4, the question how the FES is realized in concrete models is left open. 

The validity of FES was tested in two solvable models defined in the continuum space: one-dimensional bose gas interacting with the delta function and the Calogero-Sutherland model \cite{BernardWu,Murphy2,Isakov}. In those papers, it was found that the thermodynamics of the two systems are described in terms of free particles with FES. The above models, however, have no internal structures and hence the following question arises: How the FES is realized in concrete models {\it with internal structures?} 

In this paper, we establish by microscopic calculation the relation between FES and multicomponent Sutherland model (MSM). We show that the thermodynamics of MSM is the same as that of free particles with FES at all temperatures. Furthermore we obtain a simpler description with FES in low temperature region. 

This paper is organized as follows. In the next section, we review FES and introduce the statistical parameters. In section 3, we construct the thermodynamics of MSM by following the Yang-Yang's method. In section 4, we show that the MSM is thermodynamically equivalent to a free particle system with the mutual statistics at all temperatures. Furthermore, in section 5 we find that the MSM reduces to a free particle system with fractional statistics but without the mutual statistics (which is called ^^ ^^ g-on "\cite{Nayak}). 
\section{Fractional Exclusion Statistics and Thermodynamics}
General formulation and thermodynamic properties of FES were presented by several authors. \cite{Haldanefrac,Wu,Murphy} In this section, we give a brief review on the FES, in order to fix the notation and definition of terminology. 
First divide the total system into many macroscopic subsystems each of which is labeled by $\al$. Here we denote by $D_{\al}$ the number of available single particle states in the subsystem $\al$ in the presence of other particles, and by $N_{\al}$ the number of particles in the subsystem $\al$. Haldane \cite{Haldanefrac} introduced the statistical parameter $g_{\al \beta}$ through the following differential relation: 
\begin{equation}
\Delta D_{\al}=-\sum_{\beta}g_{\al \beta}\Delta N_{\beta}. \label{dDdN}
\end{equation}
Integrating eq. (\ref{dDdN}) with $N_{\al}$, we obtain
\begin{equation}
D_{\al}=G_{\al}-\sum_{\beta}g_{\al \beta}N_{\beta},\label{D}
\end{equation}
where $G_{\al}$ is the number of single particle state in the absence of other particles. 
The total number of microscopic states with $\left\{N_{\al}\right\}$ is given by
\begin{equation}
W=\prod_{\al}\frac{\left(D_{\al}+N_{\al}\right)!}{N_{\al}!D_{\al}!}\label{W}. 
\end{equation}
If we set $g_{\al \beta}=0$, equation (\ref{W}) reduces to the boson statistics in the thermodynamic limit. In the case of $g_{\al \beta}=\delta_{\al \beta}$, equation (\ref{W}) reduces to the fermion one in the thermodynamic limit. 
Here we can regard eq. (\ref{W}) as the fractional generalization of conventional quantum statistics. \cite{Wu} We term $g_{\al \beta}$ as the mutual statistical parameter when $\al\ne \beta$ and self statistical one in the case of $\al=\beta$. We call the quantum statistics governed by general values of $g_{\al \beta}$ the fractional exclusion statistics. 
Furthermore, the system is called the generalized ideal gas (GIG) if the total energy is given by
\begin{equation}
E=\sum_{\al}N_{\al}\eps_{\al}, \label{Eal}
\end{equation}
with the one particle energy $\eps_{\al}$. \cite{Wu} Here $\eps_{\al}$ is independent of the distribution $\left\{\rho_{\al}\right\}=\left\{N_{\al}/G_{\al}\right\}$. The entropy of GIG is given by taking the logarithm of eq. (\ref{W}),
\begin{eqnarray}
S&=&\sum_{\al}\left[\left(D_{\al}+N_{\al}\right)\ln\left(D_{\al}+N_{\al}\right)-D_{\al}\ln D_{\al}-N_{\al}\ln N_{\al}\right]\nonumber\\
&=&\sum_{\al}G_{\al}\left[\left(\rho_{\al}^* +\rho_{\al}\right)\ln\left(\rho_{\al}^{*}+\rho_{\al}\right)-\rho_{\al}\ln \rho_{\al}-\rho_{\al}^{*}\ln \rho_{\al}^{*}\right]\label{Sg},
\end{eqnarray}
where $\rho_{\al}^{*}=D_{\al}/G_{\al}$ is the distribution function of dual particle ( hole) of $\al$. $\rho_{\al}^{*}$ is related with $\rho_{\al}$ as 
\begin{equation}
\rho_{\al}^{*}+\sum_{\beta}\tilde g_{\al \beta}\rho_{\beta}=1\label{rhorho},
\end{equation}
with 
\begin{equation}
\tilde g_{\al \beta}\equiv g_{\al \beta}G_{\beta}/G_{\al}, 
\end{equation}
which is obtained by dividing both sides of eq. (\ref{D}) by $G_{\al}$.

Thermal average of the particle density in $\al$ is given by minimizing the thermodynamic potential $\Omega=E-TS-\sum_{\al}\mu_{\al}N_{\al}$ with respect to $\left\{\rho_{\al}\right\}$. 
The equation $\delta \Omega/\delta \rho_{\al}=0$ is expressed as
\begin{equation}
\left(1+w_{\al}\right)\prod_{\beta}\left(\frac{w_{\beta}}{1+w_{\beta}}\right)^{ g_{\beta \al}}=\exp\left[\left(\eps_{\al}-\mu_{\al}\right)/T\right], \label{wgon}
\end{equation}
with 
\begin{equation}
w_{\al}=\rho_{\al}^{*}/\rho_{\al}\label{wrhorho}.
\end{equation}
From eqs. (\ref{rhorho}) and (\ref{wrhorho}), we obtain the thermal distribution $\rho_{\al}$. 
In terms of $w_{\al}$, we can rewrite the entropy in a simple form as
\begin{equation}
S=\sum_{\al}G_{\al}\rho_{\al}\left[\left(1+w_{\al}\right)\ln \left(1+w_{\al}\right)-w_{\al}\ln \wa\right].
\end{equation}
The thermodynamic potential has also a simple form:
\begin{equation}
\Omega=-T\sum_{\al}G_{\al}\ln\left[1+\wa ^{-1}\right].\label{Omegaal}
\end{equation}

If we consider the case $g_{\al \beta}=g_{\al}\delta_{\al \beta}$, the above results are further simplified. 
The particle density is given by
 $\rho_{\al}=1/\left(w_{\al}+g_{\al}\right)$. Here $w_{\al}$ is the solution of the following equation:
\begin{equation}
\exp\left[\left(\eps_{\al}-\mu_{\al}\right)/T\right]
=
\left(w_{\al}\right)^{g_{\al}}\left(1+w_{\al}\right)^{1-g_{\al}}\label{wg}.
\end{equation}
\section{Thermodynamics of Multicomponent Sutherland Model}
Thermodynamics of multicomponent Sutherland model was constructed firstly by Sutherland and Shastry. \cite{S2} In this section, we summarize the results of thermodynamics of MSM, by following the method in ref. 9. 
We consider $N$ particle system which consists of $\KB$ component bosons and $\KF$ component fermions. The Hamiltonian is that of the modified version of the continuum Sutherland model: \cite{Poly} 
\begin{equation}
{\cal H}=
         -\frac12\sum_{i=1}^{N}\frac{\partial ^2}{\partial x_i^2}
         +\frac{\pi^2}{L^2}\sum_{i<j}
         \frac{\lambda(\lambda - M_{ij})}{\sin^2
[\pi\left(x_{i}-x_{j}\right)/L]},
\label{SP}
\end{equation}
where $x_{i}$ is the coordinate of the $i$-th particle, $L$ is the linear dimension of the system, and $\lambda$ is a coupling parameter. Then $M_{ij}$ is the coordinate exchange operator which exchanges the spatial coordinates of $i$ and $j$-th particles with the internal degrees of freedom fixed. 
 For the Hamiltonian (\ref{SP}), the energy spectrum in the thermodynamic limit has been obtained \cite{S2,Kato} as
\begin{equation}
E/L=\frac{1}{4\pi}\int_{-\infty}^{\infty}dk k^2 \nu (k)+\frac{\lambda}{8\pi}\int_{-\infty}^{\infty}dk \int_{-\infty}^{\infty}dk' \left|k-k'\right|\nu (k)\nu(k')+\frac{\pi^2 \lambda ^2 d^3}{6}, \label{Edef}
\end{equation}
where $\nu(k)$ is the momentum distribution function and $d=N/L=(1/2\pi)\int_{-\infty}^{\infty}\nu(k)dk$. Both $d$ and $\nu(k)$ have contributions from each species
\begin{equation}
d=\sum_{\alpha}d_{\alpha}=\frac{1}{2\pi}\int_{-\infty}^{\infty}dk \sum_{\al}\nu_{\al}(k), 
\end{equation}
where $\nu_{\al}(k)$ is the momentum distribution function of particle $\al$, and $d_{\al}$ is the $\al$-th component particle density. 
Since each microscopic state is identified with the momentum set occupied by particles, the entropy is given by the same form as the free particle case
\begin{equation}
S/L=\sum_{\al}s_{\al}, 
\end{equation}
with
\begin{equation}
s_{\alpha}=\frac{\pm 1}{2\pi}\int_{-\infty}^{\infty}dk 
\left[\left(1\pm \nu_{\al}\right)\ln\left(1\pm \nu_{\al}\right)\mp \nu_{\al}\ln \nu_{\al}\right]. \label{sal1}
\end{equation}
Here and in the following, the upper sign is for boson ($\al \in \B$) and the lower one for fermion ($\al \in \F$). 
Thermal equilibrium distribution function is obtained by minimizing the thermodynamic potential
\begin{equation}
\Omega=E-TS-L\sum_{\al}\mu_{\al}d_{\al}, \label{Omega}
\end{equation}
with respect to each $\nu_{\al}$. In eq. (\ref{Omega}), we have introduced the chemical potential $\mu_{\al}$ for each $\al$. 
The resultant thermal distribution functions are
\begin{equation}
\nu_{\al}(k)=\left\{\exp\left[\left(\eps(k)-\mu_{\al}\right)/T\right] \mp 1\right\}^{-1}, \label{nuB}\end{equation}
with the one particle energy 
\begin{equation}
\eps(k)=\frac{2\pi}{L}\frac{\delta E}{\delta \nu(k)}=\frac{k^2}{2}+\frac{\lambda}{2}\int_{-\infty}^{\infty}dk' \left|k-k'\right|\nu(k') +\frac{\pi^2 \lambda ^2 d^2}{2}.
\label{epsk}
\end{equation}
Equations (\ref{Omega}), (\ref{nuB}), and (\ref{epsk}) give the complete thermodynamics. For later convenience, we introduce the rapidity $p$ as
\begin{equation}
p(k)=\frac{\partial \epsilon (k)}{\partial k}
=k+\frac{\lambda}{2}\int_{-\infty}^{\infty}dk' {\rm sgn}\left(k-k'\right)\nu (k').\label{pk}
\end{equation}
Further differentiation of eq. (\ref{pk}) with $k$ gives 
\begin{equation}
\frac{\partial p(k)}{\partial k}=1+\lambda\nu (k)\label{dpdk}.
\end{equation}
By multiplying both sides of eq. (\ref{dpdk}) by $p(k)$ and integrating them over $k$, we obtain
\begin{equation}
p^2/2 =\eps\mp T\lambda\sum_{\al}\ln \left[1\pm\nu_{\al}\right]
+c_0,\label{pe}
\end{equation}
where $c_0$ is a constant which should be determined. Since $\eps \rightarrow (k+\pi \lambda d)^2/2$ and $p \rightarrow k+\pi \lambda d$ in the limit $k \rightarrow \infty$, we obtain $c_0 =0$. 
By using eq. (\ref{dpdk}) and $p(k=\pm \infty)=\pm \infty$, we can rewrite the density for $\al$ as 
\begin{equation}
d_{\al}=\frac{1}{2\pi}\int_{-\infty}^{\infty}dp\frac{\nu_{\al}}{1+\lambda \nu}.\label{dalBF}
\end{equation}
Similarly, the entropy eq. (\ref{sal1}) is rewritten as
\begin{equation}
s_{\alpha}=\frac{\pm 1}{2\pi}\int_{-\infty}^{\infty}\frac{dp}{1+ \lambda \nu} 
\left[\left(1\pm \nu_{\al}\right)\ln\left(1\pm \nu_{\al}\right)\mp \nu_{\al}\ln \nu_{\al}\right]. 
\end{equation}
The internal energy is also rewritten in terms of the integral over $p$  
\begin{equation}
E/L=\frac{1}{4\pi}\int_{-\infty}^{\infty}dp \frac{p^2 \nu}{1+\lambda \nu}\label{Ep}.
\end{equation}
Derivation of eq. (\ref{Ep}) will be given in Appendix A.
\section{Equivalence to the Generalized Ideal Gas in Thermodynamics}
In this section, we show that the thermodynamics of the MSM is equivalent to that of the free particle with FES at all temperatures. 
Firstly we introduce $\wa$ as 
\begin{equation}
\wa=\left\{
\begin{array}{cc}
1/\nu_{\al}&\mbox{for } \al \in \B \\
1/\nu_{\al}-1&\mbox{for }\al \in \F.
\end{array}
\right.\label{waBF}
\end{equation}
We rewrite eq. (\ref{pe}) with eq. (\ref{waBF}) as
\begin{equation}
p^2/2=\eps +T\lambda \sum_{\al}\ln\left[\frac{\wa}{\wa +1}\right]. \label{p2wa}
\end{equation}
By subtracting $\mu_{\al}$ from both sides of eq. (\ref{p2wa}) and taking the exponential of them, we obtain
\begin{equation}
\exp\left[\left(p^2/2 -\mu_{\al}\right)/T\right]=\left(1+\wa\right)\prod_{\beta}\left(\frac{\wb}{\wb+1}\right)^{g_{\beta \al}},\label{p2wfrac}
\end{equation}
with the ^^ ^^ statistical parameter " $g_{\beta \al}$,
\begin{equation}
g_{\beta \al}=\left\{
\begin{array}{cc}
\delta_{\al \beta}+\lambda&\mbox{for $\al \in \F$}\\
\lambda&\mbox{otherwise.}
\end{array}
\right.
\end{equation}
The quantity $\rho_{\al}= \nu_{\al}/(1+\lambda \nu)$ satisfies eq. (\ref{rhorho}) with eq. (\ref{wrhorho}).
Comparing eqs. (\ref{p2wfrac}) and (\ref{wgon}), we note that $\rho_{\al}$ is nothing but the distribution function of free particles with the statistical parameter $g_{\al \beta}$. 

Next we rewrite the thermodynamic quantities in terms of $\wa$. 
The total energy is given by $2\pi E/L=\sum_{\al}\intp \left(p^2/2\right)\rho_{\al}$. 
The entropy of species $\al$ is
\begin{eqnarray}
2\pi s_{\al}&=&\pm \int_{-\infty}^{\infty}\frac{dp}{1+\lambda \nu}\left[\left(1\pm \nu_{\al}\right)\ln \left(1+\nu_{\al}\right)\mp \nu_{\al}\ln\nu_{\al}\right]\label{salpm}\\
&=&\intp \rho_{\al}\left[\left(1+\wa\right)\ln\left(1+\wa\right)-\wa \ln \wa\right]\\
&=&\intp\left[\left(\rho_{\al}^{*}+\rho_{\al}\right)\ln\left(\rho_{\al}^* +\rho_{\al}\right)-\rho_{\al}\ln\rho_{\al}-\rho_{\al}^{*}\ln \rho_{\al}^{*}\right]. 
\label{sal}
\end{eqnarray}
By using eq. (\ref{sal}), we rewrite the thermodynamic potential as
\begin{eqnarray}
\Omega/L&=&E/L-T\sum_{\al}s_{\al}-\sum_{\al}\mu_{\al}d_{\al}\nonumber\\
        &=&\frac{1}{2\pi}\sum_{\al} \intp \rho_{\al}\left[p^2/2-\mu_{\al}-T\left(1+\wa\right)\ln\left(1+\wa\right)+T\wa \ln \wa\right]. \label{Omega1} 
\end{eqnarray}
We then use the relation (\ref{p2wfrac}) to obtain
\begin{eqnarray}
\Omega/L&=&-\frac{T}{2\pi}\sum_{\al}\intp \rho_{\al}\left[\wa\ln\left(1+\wa^{-1}\right)+\sum_{\beta} g_{\al \beta}\ln\left(1+\wb ^{-1}\right)\right]\nonumber\\
&=&-\frac{T}{2\pi}\intp \sum_{\al}\ln\left(1+\wa^{-1}\right)\sum_{\beta}\left(\delta_{\al \beta }\wa +g_{\al \beta}\right)\rho_{\beta}\nonumber\\
&=&-\frac{T}{2\pi}\intp \sum_{\beta}\ln\left(1+\wb ^{-1}\right).
\label{Omega2}
\end{eqnarray}
From the second to the third line in eq. (\ref{Omega2}), we have used eqs. (\ref{rhorho}) and (\ref{wrhorho}) with $G_{\al}=G_{\beta}$.
Compare eqs. (\ref{p2wfrac}), (\ref{sal}), and (\ref{Omega2}) with eqs. (\ref{wgon}), (\ref{Sg}), and (\ref{Omegaal}). Our results in this section show that
the thermodynamics of the MSM is equivalent to that of free fractional particle with the statistical parameter $g_{\al \beta}$ and $G_{\al}=1$ for all $\al$. 
If we set $\KB=1$ and $\KF=0$, the above results reduce to the thermodynamics of g-on with $g=\lambda$ as expected from results in ref. 5.
\section{{\normalsize g-on} description in low temperature region}

In low temperature region, we have a simpler picture of the thermodynamics of MSM. In this region, the mutual statistics can be absorbed into the self statistics and the system can be regarded as an assembly of g-ons. 
Before getting into details of calculation, first we consider the zero temperature case which will be helpful to understand how g-on emerges in low temperature region. For simplicity, consider a simple case of $\KB=1$ and $\KF=1$. At zero temperature, the distribution functions $\nu_{\B}$ for boson and $\nu_{\F}$ for fermion are given by
\begin{equation}
\nu_{\B}(k)=2\pi d_{\B}\delta \left(k\right),\quad \nu_{\F}(k)=\theta\left(k_{\F}-\left|k\right|\right),\label{nuBnuF}
\end{equation}
where $k_{\F}=\pi d_{\F}$ and $\theta (k)$ is the step function:
\begin{equation}
\theta(k)=\left\{
\begin{array}{cc}
1\quad &\mbox{for } k\ge 0 \\
0\quad &\mbox{otherwise}.
\end{array}
\right.\label{nuT0}
\end{equation}
Substitution of eq. (\ref{nuBnuF}) into eqs. (\ref{epsk}) and (\ref{pk}) gives
\begin{equation}
\eps(k)=\left\{
\begin{array}{cc}
(1+\lambda)k^2/2+\pi \lambda d_{\B}\left|k\right|+\pi^2 \lambda\left(\lambda d^2+d_{\F}^2\right)/2\quad &\mbox{ for }\left|k\right|\le k_{\F} \\
k^2/2+\pi \lambda d\left|k\right|+\pi^2 \lambda^2 d^2/2\quad &\mbox{ for } k_{\F}\le \left|k\right|,
\end{array}
\right.
\label{epsT0}
\end{equation}
\begin{equation}
p(k)=\left\{
\begin{array}{cc}
(1+\lambda)k+\pi \lambda d_{\B}\sgn(k)\quad & \mbox{ for } \left|k\right| \le k_{\F} \\
k+\pi \lambda d \sgn(k)\quad & \mbox{ for } k_{\F}\le\left|k\right|.
\end{array}
\right.
\label{pkT0}
\end{equation}
From eqs. (\ref{epsT0}) and (\ref{pkT0}), we obtain the explicit form of $\eps$ as a function of $p$
\begin{equation}
\eps=\left\{
\begin{array}{cc}
\pi^2\left(\lambda^2 d^2+\lambda d_{\F}^2\right)/2\quad&\mbox{ for }\left|p\right|\le p_{\B} \\
\left(p^2+\pi^2 \lambda \left(\lambda d +d_{\F}\right)^2\right)/2(1+\lambda)
\quad&\mbox{ for }p_{\B}\le p \le p_{\F} \\
p^2/2\quad&\mbox{ for }p_{\F}\le \left|p\right|, 
\end{array}
\right.
\end{equation}
where $p_{\B}=\pi \lambda d_{\B}$ and $p_{\F}=\pi\left(d_{\F}+\lambda d\right)$.
Figure 1 shows the $p$ dependence of $\eps$ at zero temperature. 
From eq. (\ref{dalBF}), we can regard $\rho_{\B(\F)}= \nu_{\B(\F)}/(1+\lambda \nu)$ as the rapidity distribution function of boson (fermion). In Fig. 2, the rapidity dependences of $\rho_{\B}$ and $\rho_{\F}$ at zero temperature are shown by dashed and dotted lines, respectively. \cite{note} 
The fermion distribution function $\rho_{\F}$ changes at $\left|p\right|=p_{\F}$ and $\left|p\right|=p_{\B}$ while $\rho_{\B}$ changes only at $\left|p\right|=p_{\B}$. We also show the profiles of $\rho_{\B}(p)$ and $\rho_{\F}(p)$ at low temperatures quantitatively by dotted-dotted-dashed line and dotted-dashed one, respectively. If $T$ is low enough, the temperature dependence of thermodynamic quantities comes mainly from the two regions of $p$: the vicinities of $p_{\B}$ and $p_{\F}$. The contribution to the thermodynamics from each rapidity region can be expressed in terms of thermodynamics of g-ons. The two kinds of g-ons are independent of each other, if the following condition holds:
\begin{equation}
\rho_{\B}(p=p_{\F})\simeq 0\quad\mbox{and}\quad \rho_{\F}(p=p_{\B})\simeq 1.\label{condition}
\end{equation}
The condition eq. (\ref{condition}) turns into
\begin{equation}
\DeltaBF \equiv \exp \left[\left(\muB -\muF\right)/T\right] \ll 1\label{condlow}, 
\end{equation}
where $\mu_{\B}$ and $\mu_{\F}$ are the chemical potentials of bosons and fermions, respectively. In this case, we define the ^^ ^^ low temperature region" as the region where the condition eq. (\ref{condlow}) holds. 

Now come back to the case of general value of $\KF$. In this case, however, the definition of the ^^ ^^ low temperature region " depends on the distribution of chemical potentials of each species. In the rest of this section, we consider only two cases: the {\bf case (1)} where all species of fermion have the same chemical potential and the {\bf case (2)} where all species of fermion have different chemical potentials. Generalization to other cases is straightforward. 

\subsection{the case (1)}
In this case, equation (\ref{p2wfrac}) becomes 
\begin{equation}
\exp\left[\left(p^2/2-\muB\right)/T\right]=\frac{\wB^{\lambda}}{\left(1+\wB\right)^{\lambda-1}}\left(\frac{\wF}{1+\wF}\right)^{\lambda \KF}.\label{pw11}
\end{equation}
Now we try to reduce eq. (\ref{pw11}) under the low temperature condition (\ref{condlow}). First we eliminate $\wF$ in eq. (\ref{pw11}). With the help of eq. (\ref{waBF}), we obtain the following relation:
\begin{equation}
\wF=\left(1+\wB\right)\DeltaBF.\label{wF11}
\end{equation}
Substitution of eq. (\ref{wF11}) into eq. (\ref{pw11}) gives
\begin{eqnarray}
\exp\left[\left(p^2/2-\muB\right)/T\right]&=&\frac{\wB^{\lambda}\DeltaBF ^{\lambda \KF}}{\left(1+\wB\right)^{\lambda(1-\KF)-1}\left(1+\wB \DeltaBF+\DeltaBF\right)^{\lambda \KF}}\nonumber\\
&\simeq&\frac{\wB^{\lambda}\DeltaBF ^{\lambda \KF}}{\left(1+\wB\right)^{\lambda(1-\KF)-1}\left(1+\wB \DeltaBF\right)^{\lambda \KF}}\label{substitute}.
\end{eqnarray}
Here we define the rapidity region $\Pio$ as the region where the following condition is satisfied:
\begin{equation}
\wB \DeltaBF \ll \eta,
\end{equation}
where $\eta$ is a constant which is much smaller than unity but much larger than $\DeltaBF$. In $\Pio$, the right hand side of eq. (\ref{substitute}) turns into
\begin{equation}
\wB^{\lambda}\left(1+\wB\right)^{1-\lambda(1-\KF)}\DeltaBF ^{\lambda \KF}.\label{rhs}
\end{equation}
From eqs. (\ref{substitute}) and (\ref{rhs}), we obtain
\begin{equation}
\exp\left[\left(\veps_{0}-\zeta_{0}\right)/T\right]\simeq\wB^{g_{0}}\left(1+\wB\right)^{1-g_0}\quad \mbox{ for }p\in \Pi_{0},\label{asr}
\end{equation}
where $\veps_{0}=p^2/2\Mo$, $\Mo=1+\lambda \KF$, $\go=\lambda/(1+\lambda \KF)$, and
\begin{equation}
\zeta_{0}=\muB-\lambda \KF \muF/(1+\lambda\KF). \label{zeta0} 
\end{equation}
Equation (\ref{asr}) corresponds to eq. (\ref{wg}) with energy $\veps_{0}$, chemical potential $\zeta_{0}$, and statistics $\go$. 
Here we introduce $\wo$ as the solution of the following equation:
\begin{equation}
\exp\left[\left(\veps_{0}-\zeta_{0}\right)/T\right]=\wo^{g_{0}}\left(1+\wo\right)^{1-g_0}\quad \mbox{ for all }p\label{asro}.
\end{equation}
And we rewrite the boson parts of entropy, density, and energy with $\wo$.
First consider the boson part of the entropy (eq. (\ref{salpm}) with the upper sign). Since the contribution from ${\overline \Pio}$ (the complementary region of $\Pio$) is infinitesimal (see Appendix B), we obtain
\begin{eqnarray}
2\pi s_{\B}&=&\int_{-\infty}^{\infty}\frac{dp}{1+\lambda \nu}
\left[\left(1+\nu_{\B}\right)\ln\left(1+\nu_{\B}\right)-\nu_{\B}\ln\nu_{\B}\right]\label{sBintegrand}\\
&\simeq&\int_{\Pio}\frac{dp}{1+\lambda \nu}
\left[\left(1+\nu_{\B}\right)\ln\left(1+\nu_{\B}\right)-\nu_{\B}\ln\nu_{\B}\right]\nonumber\\
&\simeq&\int_{\Pio}\frac{dp}{1+\lambda \KF+\lambda \nu_{\B}}
\left[\left(1+\nu_{\B}\right)\ln\left(1+\nu_{\B}\right)-\nu_{\B}\ln\nu_{\B}\right]\nonumber\\
&=&\Go\int_{\Pio}\frac{dp}{\wB+\go}\left[\left(1+\wB\right)\ln\left(1+\wB\right)-\wB\ln\wB\right]\nonumber\\
&\simeq&\Go\int_{\Pio}\frac{dp}{\wo+\go}\left[\left(1+\wo\right)\ln\left(1+\wo\right)-\wo\ln\wo\right],
\label{sB11}
\end{eqnarray}
with $\Go=\left(1+\lambda \KF\right)^{-1}$. It will turn out that $\Go$ is the density of states in the absence of other particles. 
For the integrand in eq. (\ref{sB11}), we can show that
\begin{equation}
\left[\left(1+\wo\right)\ln\left(1+\wo\right)-\wo \ln \wo\right]/\left(\wo+\go\right)\simeq 0\quad\mbox{ for }p\in {\overline \Pio}, 
\end{equation}
in the same way as that in Appendix B. Hence we can extend the integral region to $(-\infty,\infty)$ in the last line in eq. (\ref{sB11}) and express $2\pi s_{\B}$ as
\begin{eqnarray}
2\pi s_{\B}
&\simeq&\Go\int_{-\infty}^{\infty}\frac{dp}{\wo+\go}\left[\left(1+\wo\right)\ln\left(1+\wo\right)-\wo\ln\wo\right]\nonumber\\
&\simeq&\Go\int_{-\infty}^{\infty}dp\left[\left(\rhoo+\rhoo ^*\right)\ln\left(\rhoo+\rhoo ^*\right)-\rhoo\ln\rhoo-\rhoo ^* \ln \rhoo ^*\right]
,\label{s011}
\end{eqnarray}
with  $\rhoo=\left(\wo +\go\right)^{-1}$, which is nothing but the rapidity distribution function of g-on with the statistical parameter $\go$. Here $\rhoo ^{*}=1-\go \rhoo$ is the rapidity distribution function of dual particle (hole) of $\rhoo$. Figure 3 (a) shows the profile of $\rhoo$ and $\rhoo^{*}$ at zero and low temperatures. Both $\rhoo$ and $\rhoo ^*$ change appreciably near the generalized fermi surface $\po=\sqrt{2\Mo \zetao}$. 
Next we consider the boson density $d_{\B}$, which is rewritten as 
\begin{eqnarray}
2\pi d_{\B}&=&\intp \frac{\nu_{\B}}{1+\lambda \nu}\nonumber\\
&\simeq &\GB\int_{\Pio}\frac{dp}{\wB +\go}\nonumber\\
&\simeq &\GB\int_{\Pio}\frac{dp}{\wo +\go}\label{wogo}.
\end{eqnarray}
The integrand $\left(\wo+\go\right)^{-1}$ in eq. (\ref{wogo}) is infinitesimal since $\wo >\eta \DeltaBF^{-1}\gg 1$ in ${\overline \Pio}$. Thus we can extend the integral region from $\Pio$ to $(-\infty,\infty)$ in the last line of eq. (\ref{wogo})
\begin{eqnarray}
2\pi d_{\B}&\simeq &\GB\int_{-\infty}^{\infty}\frac{dp}{\wo +\go}\nonumber\\
&\simeq &\GB\intp\rhoo.\label{2pidB} 
\end{eqnarray}
Similarly, the boson part of the internal energy is given by
\begin{equation}
2\pi E_{\B}=\intp \frac{p^2\nu_{\B}}{2(1+\lambda \nu)}\simeq \GB\intp p^2 \rho_0/2.\label{2piEB} 
\end{equation}

Next we consider the fermion part. We define the rapidity region $\PiF$ as the region where $ \wF \geq \eta^{-1}\DeltaBF$ is satisfied. Here $\eta$ is a constant which satisfies the inequalities $\DeltaBF \ll \eta \ll 1 $. As in the boson case, the main contribution to the fermion part of entropy comes only from the region $\PiF$. In $\PiF$, we obtain
\begin{equation}
\exp\left[\left(\veps_{1}-\muF\right)/T\right]\simeq w_{\F} ^{\gl}\left(1+w_{\F}\right)^{1-\gl} \quad\mbox{ for }p\in \Pil, 
\end{equation}
with $\veps_{1}=p^2/2$ and $\gl=1+\lambda \KF$. Note that this equation is for the g-on with $\gl$ statistics. As in the boson case, we introduce $\wl$ as the solution of the following equation:
\begin{equation}
\exp\left[\left(\veps_{1}-\muF\right)/T\right]=\wl ^{\gl}\left(1+\wl\right)^{1-\gl}\quad\mbox{ for all }p. 
\end{equation}

The fermion part of the entropy has contributions only from $\PiF$ and is rewritten as 
\begin{eqnarray}
2\pi s_{\F}&=&-\KF\int_{-\infty}^{\infty}\frac{dp}{1+\lambda \nu}
\left[\left(1-\nu_{\F}\right)\ln\left(1-\nu_{\F}\right)+\nu_{\F}\ln\nu_{\F}\right]
\nonumber\\
&\simeq&-\KF\int_{\PiF}\frac{dp}{1+\lambda \KF\nu_{\F}}
\left[\left(1-\nu_{\F}\right)\ln\left(1-\nu_{\F}\right)+\nu_{\F}\ln\nu_{\F}\right]\nonumber\\
&\simeq&\GF\int_{\Pil}\frac{dp}{\wF+\gl}\left[\left(1+\wF\right)\ln\left(1+\wF \right)-\wF\ln\wF\right]\nonumber\\
&\simeq&\GF\int_{-\infty}^{\infty}\frac{dp}{\wl+\gl}\left[\left(1+\wl\right)\ln\left(1+\wl\right)-\wl\ln\wl\right]\nonumber\\
&\simeq&\GF\intp\left[\left(\rhol+\rhol^{*}\right)\ln\left(\rhol+\rhol^{*}\right)-\rhol\ln\rhol-\rhol^{*}\ln\rhol^{*}\right].\label{sF11}
\end{eqnarray}
Here $\rhol=\left(\wl+\gl\right)^{-1}$ and $\rhol^{*}=1-\gl \rhol$. The profiles of $\rhol$ and $\rhol ^*$ are  shown in Fig. 3(b). The generalized fermi surface of $\rhol$ is given by $p_{1}=\sqrt{2 \mu_{\F}}$. 

In contrast to the case of entropy, the fermion parts of density and energy have contributions from $\PiB$ as well as from $\PiF$. 
The fermion part of density is
\begin{eqnarray}
2\pi d_{\F}&=      &\KF\intp\frac{\nu_{\F}}{1+\lambda \nu}\nonumber\\
          &\simeq &\int_{\Pil}dp \frac{\KF\nu_{\F}}{1+\lambda \KF\nu_{\F}}                       -  \KF  \int_{\Pio}dp \left\{\frac{1}{1+\lambda \KF}-\frac{1}{1+\lambda \KF+\lambda \nu_{\B}}
                             \right\}\nonumber\\
&\simeq&\KF\intp \left(\rhol-\lambda \GB^2 \rhoo\right)\label{2pidF}.
\end{eqnarray}
From eqs. (\ref{2pidB}) and (\ref{2pidF}), we obtain 
\begin{equation}
2\pi \left(\muB d_{\B}+\muF d_{\F}\right)\simeq\intp\left(\GB \zetao\rhoo+\GF \muF\rhol\right).\label{mud}
\end{equation}
Similarly, the total energy is given by
\begin{equation}
2\pi E/L\simeq\intp\left(\GB \veps_{0}\rhoo+\GF \veps_{1}\rhol\right).\label{E11} 
\end{equation}
From eqs. (\ref{s011}), (\ref{sF11}), (\ref{mud}), and (\ref{E11}), we obtain the thermodynamic potential as
\begin{equation}
2\pi\Omega/L\simeq-T\intp\left[\GB\ln\left(1+\wo ^{-1}\right)+\Gl\ln\left(1+\wl^{-1}\right)\right].\label{Omega11}
\end{equation}
Equation (\ref{Omega11}) is obtained in the same way as that in the derivation of eq. (\ref{Omega2}). 
We remark that $\KF$ fermi components have the same $G_{\al}=1$ as in the multicomponent FES description, but that the boson component has $G_{0}<1$ in the g-on description.  
From eq. (\ref{Omega11}), we find that the thermodynamics of MSM is described in terms of two kinds of g-ons independent of each other. 
\subsection{case (2)}
Next consider the case where each species of fermion has a chemical potential different from one another; $\mu_{\F,1}>\mu_{\F,2}>\cdots >\mu_{\F,\KF}>\mu_{\B}$. Here $\mu_{\F,q}$ is the chemical potential of $q$-th component of fermion. In this case, we define the low temperature region as the region where the following inequalities are satisfied:
\begin{equation}
\exp\left[\left(\muB-\mu_{\F,\KF}\right)/T\right]\ll 1, 
\end{equation}
\begin{equation}
\exp\left[\left(\mu_{\F,q+1}-\mu_{\F,q}\right)/T\right]\ll 1\quad \mbox{ for }q=1 \sim \KF-1.\end{equation}
In this case, equation (\ref{p2wfrac}) becomes
\begin{equation}
\exp\left[\left(p^2/2-\muB\right)/T\right]=\frac{\wB^{\lambda}}{\left(1+\wB\right)^{\lambda-1}}\prod_{q=1}^{\KF}\left(\frac{\wFq}{1+\wFq}\right)^{\lambda}, \label{pwq11}
\end{equation}
where  $\wFq= \left[\nu_{\F}\left(\eps -\mu_{\F,q}\right)\right]^{-1}-1$ and $\wB=\nu_{\B}(\veps-\mu_{0})^{-1}$. From the similar study with the previous case, we find that the boson contributions in the present case to the thermodynamic quantities are given by the same expressions as those in the previous case, after replacing the expression (\ref{zeta0}) by $\zetao=\muB-\lambda \sum_{q=1}^{\KF}\mu_{\F,q}/(1+\lambda \KF)$. 

Now we consider the $q$-th component fermion part. First rewrite the above equation (\ref{pwq11}) in terms of
 $\wFq$ in the rapidity region $\Piq$ where $\wFq \ll {\rm min}[\Delta_{r,q}]$ for $1\le r <q \le \KF$. Here $\Delta_{r,q}\equiv \exp\left[\left(\mu_{\F,r}-\mu_{\F,q}\right)/T\right]$. In this rapidity region, we have
\begin{equation}
\frac{\wFr}{1+\wFr}=\frac{\wFq \Delta_{q,r}}{1+\wFq \Delta_{q,r}}\simeq\left\{
\begin{array}{cl}
\wFq \Delta_{q,r}\quad&\mbox{ for }r <q\\
1\quad&\mbox{ for } q<r, 
\end{array}
\right.
\label{replace11}
\end{equation}
\begin{equation}
\left(\frac{\wB}{1+\wB}\right)=\frac{\wFq \Delta_{q,\B}-1}{\wFq \Delta_{q,\B}}\simeq 1. \label{wBreplace11}
\end{equation}
Here $\Delta_{q,\B}\equiv\exp\left[\left(\mu_{\F,q}-\muB\right)/T\right]$. 
Using eqs. (\ref{wBreplace11}) and (\ref{replace11}), equation (\ref{pwq11}) becomes
\begin{equation}
\exp\left[\left(\eps_q -\zeta_q\right)/T\right]\simeq\wFq^{g_q}\left(1+\wFq\right)^{1-g_q}\quad\mbox{ for }p\in \Piq\label{wFq}, 
\end{equation} 
with $\eps_q=p^2/(2M_q)$, $M_{q}=1+\lambda \left(q-1\right)$, $\zeta_q=\mu_{\F,q}-\lambda \Gq \sum_{r=1}^{q-1}\mu_{\F,r}$, $g_q=1+\lambda \Gq$, and
\begin{equation}
\Gq=M_{q}^{-1}=\left(1+\lambda\left(q-1\right)\right)^{-1}. \label{Gq} 
\end{equation}
Equation (\ref{wFq}) corresponds to eq. (\ref{wg}) with the energy $\eps_q$, chemical potential $\zeta_q$ and statistical parameter $g_q$. At the end of this section, we will find that $\Gq$ in eq. (\ref{Gq}) is the density of states of $g_q$-on in the absence of other particles. 

We define $\wq$ ($1\le q \le \KF$) as the solution of
\begin{equation}
\exp\left[\left(\eps_q -\zeta_q\right)/T\right]=\wq^{g_q}\left(1+\wq\right)^{1-g_q}\quad\mbox{ for all }p.\label{wq}
\end{equation} 
Now we rewrite the entropy, density, and energy in terms of $\wq$. 
The contribution to the entropy from $q$-th component is given by
\begin{eqnarray}
2\pi s_{q}&=&-\int_{-\infty}^{\infty}\frac{dp}{1+\lambda \nu}\left[\left(1-\nu_{q}^{\F}\right)\ln\left(1-\nu_{q}^{\F}\right)+\nu_{q}^{\F}\ln\nu_{q}^{\F}\right]\nonumber\\
&\simeq&G_q\intp\left[\left(\rho_q+\rho_q^{*}\right)\ln\left(\rho_q+\rho_q^*\right)-\rho_q\ln\rho_q-\rho_q^*\ln\rho_q^*\right]\label{sq}.
\end{eqnarray}
Here $\rhoq =\left(\wq+\gq\right)^{-1}$ and $\rho_q^*= 1-g_q \rho_q$ are the distribution functions of $\gq$-on and its hole, respectively. 

The $q$-th component fermion density is given by
\begin{equation}
2\pi d_q=\intp\frac{\nu_{q}^{\F}}{1+\lambda \nu}.\label{dq}
\end{equation}
Figure 4 shows the profile of the integrand in eq. (\ref{dq}), $\nu_q (p)=\nu_{q}^{\F}/(1+\lambda \nu)$, at zero and low temperature in the case where $\KF=2$ and $q=1$. The generalized fermi surfaces are $p_{q}=\sqrt{2M_q \zeta_q}\quad (q=0,1,2)$. With the help of Fig. 4, we see that $d_q$ has a positive contribution from the region $\Piq$ and negative ones from $\Pir$ $(r > q \mbox{ or }r=0)$ 
\begin{eqnarray}
2\pi d_q&\simeq&
\Gq\int_{\Piq} \frac{dp}{\wq +\gq}-\lambda \sum_{r=q+1}^{\KF}\Gr ^2\int_{\Pir} \frac{dp}{\wrkato +\gr}-\lambda \Go ^2\int_{\Pio}\frac{dp}{\wo +\go}\nonumber\\
&\simeq&\Gq\intp \rhoq-\lambda \sum_{r=q+1}^{\KF}\Gr^2\intp \rhor-\lambda \Go^2\intp \rhoo.
\label{dq2} 
\end{eqnarray}
Similarly, the energy of $q$-th component fermion is given by
\begin{eqnarray}
2\pi E_{q}&= &\intp\frac{p^2\nu_{q}^{\F}}{2(1+\lambda \nu)}\\
&\simeq&\Gq\intp \frac{p^2\rho_q}{2}-\lambda \sum_{r=q+1}^{\KF}\Gr^2\intp \frac{p^2\rho_{q+r}}{2}-\lambda \Go^2 \intp \frac{p^2\rhoB}{2}\label{dq3}. 
\end{eqnarray}
From eq. (\ref{dq3}), we obtain the following result:
\begin{equation}
2\pi \left(\muB d_{\B}+\sum_{q=1}^{\KF}\mu_{\F,q} d_{q}\right)
\simeq\sum_{q=0}^{\KF}\zeta_q \Gq\intp \rho_q.\label{mud2}
\end{equation}
The total energy is given in a form similar to eq. (\ref{mud2}),
\begin{equation}
2\pi E/L\simeq\sum_{q=0}^{\KF}\Gq\intp\eps_q \rho_q.\label{Erho}
\end{equation}
From eqs. (\ref{wq}), (\ref{sq}), (\ref{mud2}), and (\ref{Erho}), the thermodynamic potential in the low temperature region reduces to
\begin{equation}
2\pi\Omega/L\simeq-T\sum_{q=0}^{\KF}\Gq\intp\ln\left(1+w_q^{-1}\right). 
\end{equation}
\section{Discussion}
In the last section, we obtain the g-on description in the low temperature regime for the two cases ( (1) and (2) ). Our results imply that the presence of a ^^ ^^ magnetic field", which polarizes the system, leads to a g-on description different from that obtained in the absence of the magnetic field. As we increase the magnetic field with $T$ fixed, thermodynamics at low temperature changes from the one described by the case (1) to another obtained in the case (2). 
Such a statistical crossover of g-on occurs only in the multicomponent model. 

In section 4, we proved that the thermodynamics of MSM is The same as that of particles obeying FES with certain statistical parameters. Here we discuss the validity of FES description to other solvable models; recently we proved that the thermodynamics of the Haldane-Shastry and long-ranged {\it t-J} models are described exactly in terms of the thermodynamics of free particle with FES at all temperatures. \cite{KatotJ}
 Thermodynamics of the Bethe solvable bose gas is proved to be equivalent with that of particles obeying the FES in ref. 5. 
For the isotropic Heisenberg model with the nearest neighbor exchange, where the string solutions exist, thermodynamics is described in terms of FES after identifying the $n$ string with $n$-th species (see Appendix C). From the above results, we expect that thermodynamics of other Bethe solvable models are interpreted in terms of FES. 
However, we have not yet obtained the exact FES description of the thermodynamics of the multicomponent Haldane-Shastry and long-ranged $t-J$ models. Hence we are not sure that thermodynamics of {\it all} solvable models are exactly described in terms of FES at the present stage.

In section 5, we found that g-on description accounts for the low temperature thermodynamics of MSM. On the other hand, the validity of g-on description to the lattice versions of the Sutherland model is limited; we found that the g-on descriptions provide the correct low temperature properties of Haldane-Shastry, long-ranged {\it t-J}, and its multicomponent models in the presence of the magnetic field. In the absence of the magnetic field, however, thermodynamics of the models are described in terms of the parafermion statistics \cite{KuramotoKato}, which cannnot be described in terms of g-on. Though we have not yet studied the Bethe solvable models at this point, we expect the low temperature properties of those models would be described in terms of a picture different from the g-on one. 
\section*{Acknowledgement}
We acknowledge the support by a Grant-in-Aid from the Ministry of Education, Science, and Culture, Japan. 
\appendix
\section*{A}
In this appendix, we derive eq. (\ref{Ep}) from eq. (\ref{Edef}).
By using eq. (\ref{pk}), we obtain,
\begin{eqnarray}
\intk p^2\nu(k)&=&\intk k^2 \nu +\lambda \intk k\nu(k)\intkp \nu(k_1)\sgn(k-k_1)\nonumber\\
&+&\frac{\lambda^2}{4}\intk \nu(k)\intkp \nu(k_1)\sgn(k-k_1)\intkpp \nu(k_2)\sgn (k-k_2)\label{p2}.
\end{eqnarray}
The second term in eq. (\ref{p2}) is rewritten as
\begin{eqnarray}
& &\frac{\lambda}{2}\intk\intkp \left(k-k_1\right)\sgn\left(k-k_1\right)\nu(k)\nu(k_1)\nonumber\\
&=&\frac{\lambda}{2}\intk\intkp\left|k-k_1\right|\nu(k)\nu(k_1).\label{rhs2}
\end{eqnarray}
The third term in eq. (\ref{p2}) is rewritten as
\begin{eqnarray}
& &\frac{\lambda^2}{2}\intk \intkp\int_{-\infty}^{k_1} dk_2\nu(k)\nu(k_1)\nu(k_2)\sgn(k-k_1)\sgn(k-k_2)\nonumber\\&=&\frac{\lambda^2}{2}\intk \int_{-\infty}^{k}dk_1\int_{-\infty}^{k_1}dk_2\nu(k)\nu(k_1)\nu(k_2)\nonumber\\
&=&\frac{\lambda^2}{12}\intk \nu(k)\intkp \nu(k_1)\intkpp \nu(k_2)\nonumber\\
&=&\frac{\left(2\pi d\right)^3 \lambda^2}{12}=\frac{2\pi^3 \lambda ^2 d^3}{3}\label{rhs3}
\end{eqnarray}
From eqs. (\ref{p2}), (\ref{rhs2}), and (\ref{rhs3}), we find that
\begin{equation}
2\pi E/L=\intk \frac{p^2 \nu (k)}{2}.
\end{equation}
Replacing $\intk \rightarrow \int_{-\infty}^{\infty}dp/(1+\lambda \nu)$, we obtain eq. (\ref{Ep}).
\section*{B}
In this Appendix, we show that the contribution from the rapidity region ${\overline \Pio}$ to $2\pi s_{\B}$ is negligible under the condition (\ref{condlow}) in the case (1). In the rapidity region ${\overline \Pio}$, $\wB \geq \eta \DeltaFB$ holds. The numerator in the integrand in eq. (\ref{sBintegrand}) is rewritten as
\begin{eqnarray}
& &\left[\left(1+\nuB\right)\ln\left(1+\nuB\right)-\nuB \ln \nuB\right]\nonumber\\
&=&\left[\left(1+\wB^{-1}\right)\ln\left(1+\wB^{-1}\right)-\wB^{-1} \ln \wB^{-1}\right]\simeq {\cal O}(\wB^{-1})\leq {\cal O}((\DeltaFB\eta)^{-1})\ll 1,
\end{eqnarray}
while the denominator $1+\lambda \nu$ is larger than unity. 
Hence the integrand in eq. (\ref{sBintegrand}) is infinitesimal in the region ${\overline \Pio}$.
\section*{C}
Here we consider the one-dimensional isotropic Heisenberg model under the periodic boundary condition:
\begin{equation}
{\cal H}=\frac14 \sum_{i=1}^{N}\left(\mbox{{\boldmath$\sigma$}}_{i}\cdot \mbox{{\boldmath$\sigma$}}_{i+1}-1\right)+H\sum_{i=1}^{N}\sigma _{i}^{z}, 
\label{XXX}\end{equation}
$$
\mbox{{\boldmath$\sigma$}}_{1}=\mbox{{\boldmath $\sigma$}}_{N+1}.
$$
Here $\mbox{{\boldmath$\sigma$}}_{i}$ is the Pauli matrix at the site $i$ and $N$ is the number of site. Magnetic field $H$ is applied parallel to $z$-axis. 
The thermodynamics of the model eq. (\ref{XXX}) were first constructed by Takahashi \cite{Takahashi} by using the string hypothesis. In ref. 15, the thermodynamics is given as follows.  
The internal energy $E$ is
\begin{equation}
E/N=\sum_{n=1}^{\infty}\int_{-\infty}^{\infty}dk \veps_{0}^{n}(k)\nu_{n}(k)-H,
\end{equation}
with 
\begin{equation}
\veps_{0}^{n}(k)=\frac{-2n}{k^2+n^2}+2nH, 
\end{equation}
where $k$ is the rapidity and $\nu_{n}(k)$ is the rapidity distribution function of the string with the length $n$. The entropy $S$ is given by
\begin{equation}
S/N=\sum_{n=1}^{\infty}\int_{-\infty}^{\infty}dk \left[\left(\nu_{n}+\nu^{*}_{n}\right)\ln\left(\nu_{n}+\nu^*_n\right)-\nu_{n}\ln \nu_n -\nu_n ^{*}\ln \nu_{n}^{*}\right].
\end{equation}
Here $\nu_{n}^{*}$ is the rapidity distribution function of hole of n-string. Among $\nu^*_{n}$ and $\nu_m$, there is the following relation:
\begin{equation}
G_{n}(k)=\nu^{*}_{n}(k)+\sum_{m=1}^{\infty}\int_{-\infty}^{\infty}dk'g_{nm}(k-k')\nu_m (k'),\label{relation}
\end{equation}
where $G_{n}(k)=n/\pi(n^2+k^2)$ and $g_{nm}(k)$ is given by
\begin{equation}
g_{nm}(k)=f_{\left|n-m\right|}(k)+2f_{\left|n-m\right|+2}(k)+2f_{\left|n-m\right|+4}(k)+\cdots +2f_{n+m-2}(k)+f_{n+m}(k),
\end{equation}
where $f_{n}(k)=n/\pi (n^2+k^2)$, for $n>0$ and $f_{0}(k)=\delta(k)$. 
Here we introduce $\rho_{n}=\nu_{n}(k)/G_{n}(k)$ and $\rho^*_n(k)=\nu^*_n (k)/G_{n}(k)$. In terms of $\rho_n$ and $\rho_n ^*$, the relation (\ref{relation}) becomes
\begin{equation}
1=\rho_n^*+\sum_{m=1}^{\infty}\int_{-\infty}^{\infty}dk'\tilde g_{nm}(k-k')\rho_{m}(k'),\label{relation2}
\end{equation}
with $\tilde g_{nm}=g_{nm}G_{m}/G_n$.
Similarly, the energy and entropy are given as
\begin{equation}
E/N=\sum_{n=1}^{\infty}\int_{-\infty}^{\infty}dkG_{n}\veps_{n}^0\rho_n
\label{energy2}\end{equation}
\begin{equation}
S/N=\sum_{n=1}^{\infty}\int_{-\infty}^{\infty}dk G_{n}\left[\left(\rho_n+\rho_n^*\right)\ln \left(\rho_n +\rho_n^*\right)-\rho_n \ln \rho_n-\rho_n^* \ln \rho_{n}^*\right].\label{entropy2}
\end{equation}
From the expressions (\ref{relation2}), (\ref{energy2}), and (\ref{entropy2}), we can regard the thermodynamics of isotropic Heisenberg model as that of GIG with energy $\veps_n^0$, statistical parameter $g_{nm}(k)$, and the density of states $G_{n}$. 

\begin{figure}
\caption{Rapidity dependence of one particle energy \protect$\eps(p)$. Two dots represent the points where the derivative with \protect$p$ is discontinuous.}
\end{figure}
\begin{figure}
\caption{Rapidity distribution functions of boson $\rhoB(p)$ and fermion $\rhoF (p)$ are shown in the case where $\KB=\KF=1$. The dashed and the dotted lines represent $\rhoB (p)$ and $\rhoF (p)$ at zero temperature, respectively. The dotted-dotted-dashed line and dotted-dashed one represent $\rhoB (p)$ and $\rhoF (p)$ respectively at low temperature.}
\end{figure}
\begin{figure}
\caption{ (a) Rapidity dependences of $\rhoo (p)$ and $\rhoo ^{*}(p)$ at low temperature are shown by dashed and dotted lines, respectively. (b) Rapidity dependences of $\rhol (p)$ and $\rhol ^{*}(p)$ at low temperature are shown. In both figures, results at zero temperature are also shown, for comparison. Note that $\rhol (p)$ has no structure at $p=p_{0}$ in contrast to $\rho_{\F}(p)$ in Fig. 2.}
\end{figure}
\begin{figure}
\caption{The profile of the integrand in eq. (\protect\ref{dq}). $\nu_{1}(p)$ in the case of $\KF=2$ is shown at low temperature. The generalized fermi surfaces for $\rhoo$, $\rho_{1}$, and $\rho_{2}$ are represented by $p_{0}$, $p_{1}$, and $p_{2}$, respectively. For a reference, the result at zero temperature is also shown by dotted line. }
\end{figure}

\begin{references}
\bibitem{note}It should be noted that the boson does not turn into g-on with $g=1/\lambda$. In order to determine the correct value of $g$, we must consider the normalization factor.
\end{references}
\end{document}